# Thermal Stability of Hafnium Zirconium Oxide on Transition Metal Dichalcogenides

*Maria Gabriela Sales[1], Samantha T. Jaszewski[1], Shelby S. Fields[1], Jon F. Ihlefeld[1,2], Stephen J. McDonnell[1,\*]*

[1]Department of Materials Science and Engineering, University of Virginia, Charlottesville, VA 22904

[2]Department of Electrical and Computer Engineering, University of Virginia, Charlottesville, VA 22904

*Corresponding author: mcdonnell@virginia.edu

Keywords: Hafnium zirconium oxide, $MoS_2$, $WSe_2$, atomic layer deposition, X-ray photoelectron spectroscopy

Abstract

Ferroelectric oxides interfaced with transition metal dichalcogenides (TMDCs) offer a promising route toward ferroelectric-based devices due to lack of dangling bonds on the TMDC surface leading to a high-quality and abrupt ferroelectric/TMDC interface. In this work, the thermal stability of this interface is explored by first depositing hafnium zirconium oxide (HZO) directly on geological $MoS_2$ and as-grown $WSe_2$, followed by sequential annealing in ultra-high vacuum (UHV) over a range of temperatures (400-800 °C), and examining the interface through X-ray photoelectron spectroscopy (XPS). We show that the nucleation and stability of HZO grown through atomic layer deposition (ALD) varied depending on functionalization of the TMDC, and the deposition conditions can cause tungsten oxidation in $WSe_2$. It was observed that HZO deposited on non-functionalized $MoS_2$ was unstable and volatile upon annealing, while HZO on functionalized $MoS_2$ was stable in the 400-800 °C range. The $HZO/WSe_2$ interface was stable until 700 °C, after which Se began to evolve from the $WSe_2$. In addition, there is evidence of oxygen vacancies in the HZO film being passivated at high temperatures. Lastly, X-ray diffraction (XRD) was used to confirm crystallization of the HZO within the temperature range studied.

Introduction

Ferroelectrics have been utilized in non-volatile memory devices owing to their bi-stable and reversible polarization states and low power consumption [1]. The most common ferroelectrics in such devices, such as lead zirconate titanate, however, suffer from size scaling effects [2] that limit the accessible technology node to 130 nm [3]. The relatively recent discovery of ferroelectricity in sub-20 nm thick, crystalline silicon-doped $HfO_2$ [4] offers a potential pathway toward continued device scaling [5] and development of new devices utilizing the switchable polarization. Such devices include ferroelectric field effect transistors (FeFETs) [6, 7] and negative differential capacitance FETs [8, 9]. The ferroelectric response in $HfO_2$-based materials has been attributed to a metastable non-centrosymmetric orthorhombic phase, with space group $Pca2_1$ [10]. This orthorhombic phase may be stabilized over the equilibrium centrosymmetric monoclinic phase through reduced film thickness and/or constrained grain size [11, 12], biaxial stress [13-15], oxygen vacancies [16, 17], and dopants, such as Si, Al, Y, Gd, and Zr [4, 18-21]. Zr substitution for Hf to form hafnia zirconia alloys is particularly promising for device applications, as these alloys exhibit low crystallization temperatures [22], possess ferroelectric properties that are resistant to degradation with annealing [23], and demonstrate a large composition space that enable ferroelectric phase formation [19].

Some of the potential ferroelectric-based memory devices, such as FeFETs, require contact between the ferroelectric oxide, acting as a gate dielectric, and semiconducting channel [24-28]. A poor interface between ferroelectric materials and silicon presents a challenge in integration with traditional semiconductor technology. Some issues include interdiffusion of constituents across the interface, and the presence of an interfacial dielectric layer ($SiO_2$) between the ferroelectric and silicon [29, 30]. Thus, a potential alternative would be to use transition metal dichalcogenides (TMDCs) as the semiconducting material in FeFETs. Due to their 2D van der Waals nature, TMDCs have strong in-plane covalent bonds with no out-of-plane dangling bonds, thus providing high in-plane chemical stability, leading to an improved interface quality with the ferroelectric without intermixing or interfacial chemical species [31]. In this work, we consider the thermal stability of this interface, which is important for identifying potential processing limitations.

Hafnium zirconium oxide (HZO) films deposited by atomic layer deposition (ALD) are typically amorphous and require annealing at temperatures ranging from 400-800 °C for crystallization [14, 15, 32]. Thus, the ferroelectric/TMDC heterostructure must be able to withstand these temperature excursions. Exposing ferroelectric/TMDC heterostructures to high temperatures in the fabrication of ferroelectric-based devices has been shown in previous work [26], with the focus being mainly on the device fabrication and not on the interface itself. The present study is limited to the direct interface between the ferroelectric and the TMDC.

In this study, we explore HZO as our ferroelectric, and examine its interface with semiconducting TMDCs, $MoS_2$ and $WSe_2$. While it is conventional to prepare HZO on an electrode (*e.g.* TiN) [19], doing so would require transfer of the TMDC onto the HZO after the HZO is prepared. It is known that transfer processes introduce contaminants by leaving behind residues that could affect the interface and device properties [33-37]. Thus, in this work, we directly grow HZO on TMDCs through ALD, and show differences in the efficacy of the ALD process depending on the TMDC and its functionalization. The focus of this work is to study how the interface changes at anneal temperatures that are conventional to crystallizing HZO films (*i.e.* 400-800 °C).

Methodology

Two types of geological MoS$_2$ (SPI Supplies [38]) substrates were used – untreated and ultraviolet-ozone (UV-O$_3$) treated. The untreated sample was mechanically exfoliated with Scotch tape in order to remove the topmost layers of MoS$_2$ and expose a clean surface. The UV-O$_3$ treated sample was a similarly exfoliated sample that underwent UV-O$_3$ treatment in air for 30 s using a UV grid lamp connected to a 3 kV, 0.03 A power supply (BHK, Inc.). Prior work has shown that UV-O$_3$ enhances ALD nucleation on MoS$_2$ and leads to more uniform growth of Al$_2$O$_3$ and HfO$_2$ [39, 40]. Previous reports have used UV-O$_3$ treated MoS$_2$ in transistors [41, 42], and have shown that this UV-O$_3$ functionalization of MoS$_2$ allowed deposition of a continuous HfO$_2$ dielectric layer with low leakage [43].

WSe$_2$ was grown on a 1 x 1 cm square of highly ordered pyrolytic graphite (HOPG) from SPI Supplies [38] using molecular beam epitaxy (MBE) in an ultra-high vacuum (UHV) system described elsewhere [44]. The HOPG crystal was outgassed for approximately 12 h in the MBE chamber at 250 °C. The WSe$_2$ growth was performed at 550 °C with a Se:W flux ratio of 7500:1. An initial ripening step was carried out wherein the sample was periodically exposed to the W flux for 30 s in 90 s intervals while the Se flux was kept constant. This interrupted growth technique is adapted from literature [45, 46]. Additionally, the use of a ripening step has also been demonstrated in other reports [47, 48]. The ripening step was conducted for a total of 38 min, after which the sample was cooled in a Se flux until it reached 375 °C. At this temperature, the Se flux was terminated, and the sample was continuously cooled to below 100 °C. The sample was then reheated to growth temperature. An uninterrupted growth followed, in which the W and Se fluxes were kept constant for 2 h 40 min. The growth rate was monitored through in-situ reflective high energy electron diffraction (RHEED), and the growth was ended when approximately 4 layers of WSe$_2$ were formed on the substrate. At the end of the growth, the sample was cooled to 275 °C in a Se flux to minimize chalcogen vacancies. At 275 °C the Se flux was turned off to avoid Se adsorption on the surface, and the sample was further cooled to room temperature. This employed MBE growth procedure results in repeatable WSe$_2$ chemistry, as will be discussed in future publications.

The untreated MoS$_2$, UV-O$_3$ treated MoS$_2$, and as-grown WSe$_2$ were all loaded side-by-side into an Oxford FlexAL II ALD instrument for deposition of HZO. ALD was conducted at 150 °C utilizing Tetrakis(ethylmethylamido)hafnium (TEMA-Hf) and tetrakis(ethylmethylamido)zirconium (TEMA-Zr) precursors in a 1:1 dose ratio with H$_2$O as an oxidant and argon as a carrier gas. The precursor temperatures were 70 °C for TEMA-Hf and 85 °C for TEMA-Zr. For HZO deposition, unless otherwise stated, the deposition sequence was such that the TEMA-Hf precursor was introduced first, followed by oxidant, and then the TEMA-Zr precursor, followed by oxidant. The pulse and purge times were 1 s and 3 s, respectively, for TEMA-Hf, and 1.5 s and 6 s, respectively, for TEMA-Zr. Following each precursor dose, the H$_2$O oxidant was pulsed for 20 ms and evacuated for 16 s. The growth rates per cycle for HfO$_2$ and ZrO$_2$, calibrated on bare silicon substrates, were approximately 1.2 Å/cycle and 1.4 Å/cycle, respectively. In addition, we note that these ALD conditions are optimized for uniformity across a 100 mm wafer, and the three samples loaded were all placed within this 100 mm diameter. A film thickness of approximately 1 nm (4 total Hf and Zr cycles) was targeted for samples for interface chemistry studies using X-ray photoelectron spectroscopy (XPS). The 1 nm film thickness allows measurement of the XPS core levels of both the underlying TMDC substrate as well as the HZO film on the surface.

Sequential heating and XPS were performed on the HZO/TMDC samples in the same UHV chamber described previously [44]. XPS spectra were taken using a monochromated Al Kα X-ray source at 300 W with a Scienta Omicron R3000 analyzer at a pass energy of 50 eV. XPS measurements were acquired after HZO deposition, after annealing in UHV for 20 min at 400 °C, and after annealing in UHV for 40 min each at 500 °C, 600 °C, 700 °C, and 800 °C. The anneals at each temperature were performed on all three samples simultaneously, meaning that the three samples were maintained at the same temperature for the same amount of time for each annealing step. Spectral analysis of XPS results was carried out using kolXPD software [49]. All peaks were fit with Voigt lineshapes.

In addition to the samples for XPS studies, two separate samples with thicker layers of HZO were prepared for X-ray diffraction (XRD) measurements to assess crystallization. Approximately 10 layers of WSe$_2$ was MBE-grown on an AlN/Si (111) substrate using the same method described for the WSe$_2$ growth on HOPG. After MBE growth, ~20 nm thick HZO (83 total Hf and Zr cycles) was deposited on the WSe$_2$ by ALD under the same conditions as described above. Note that for these samples for XRD, AlN/Si (111) was used as the substrate instead of HOPG because the (002) peak of HOPG at 26.6° in 2θ would interfere with the signal of HZO, whose peaks of interest are within the 2θ range of 26-33°. A Rigaku Smartlab X-ray Diffractometer using Cu Kα radiation in a parallel beam configuration with a fixed incident angle of 0.7° (greater than the critical angle for total external reflection) was used for grazing-incidence X-ray diffraction (GIXRD) measurements to assess crystallinity and phase assemblage. GIXRD measurements were performed in the 2θ range of 26-33° as this region possesses reflections of the possible polymorphs: the monoclinic *P*1 2$_1$/*c*1 phase, tetragonal *P*4$_2$/*nmc* phase, and orthorhombic *Pca*2$_1$ phase.

Results and Discussion

HZO Atomic Layer Deposition Efficacy

Figure 1 shows the XPS spectra of the TMDC substrates immediately following HZO deposition. The detected features correspond to the expected peaks based on the chemical bonding environment of $MoS_2$ and $WSe_2$ (Mo-S and W-Se bonds, respectively). We note that for the $WSe_2$ sample, oxidation of W (W-O bonds) from the ALD process was evident, and this is discussed further in later sections. A non-covalent interface between the TMDCs and HZO film is expected, as a smooth interface without intermixing between TMDCs and oxides such as $HfO_2$ [40, 50-52], $ZrO_2$ [53], and $Al_2O_3$ [39, 52, 53] have been observed through photoemission spectroscopy and electron microscopy techniques in previous reports.

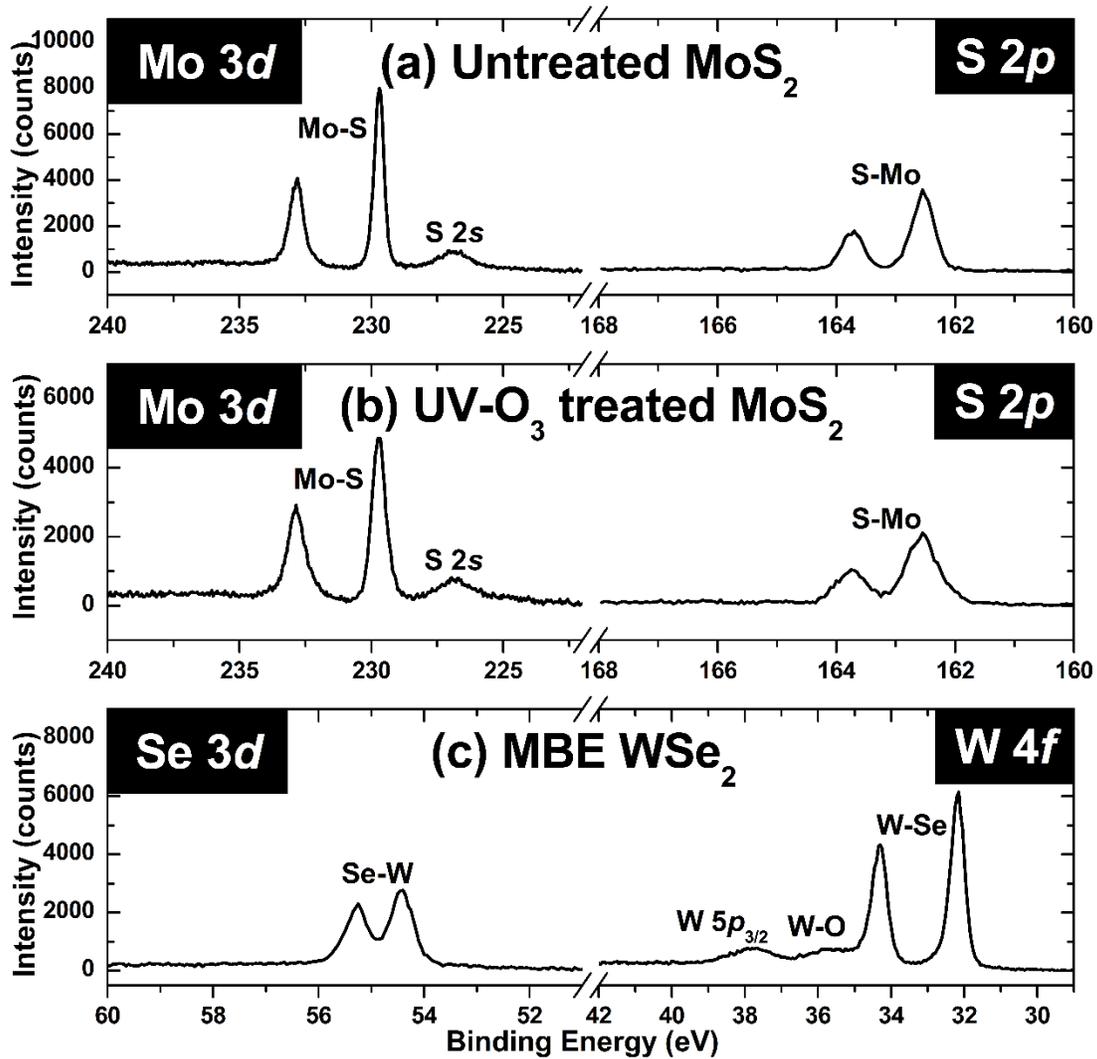

**Figure 1.** XPS spectra of the TMDC substrates after HZO deposition: untreated $MoS_2$ (a), UV-$O_3$ treated $MoS_2$ (b), and $WSe_2$ grown via MBE (c). The different core levels and chemical states are labeled.

Shown in Fig. 2 is the Hf 4$f$ region for all three samples after HZO deposition. The peak intensities are related to the amount of HZO deposited on each sample. It should be noted that the ALD conditions used were optimized for uniformity across a 100 mm wafer, indicating that any differences in thickness are not attributable to ALD processing nonuniformities. Since HZO deposition was performed with all three samples in parallel, and all were exposed to air for the same amount of time between the ALD process and XPS characterization, the differences in peak intensities can be confidently assumed to indicate differences in the efficacy of the same HZO deposition process.

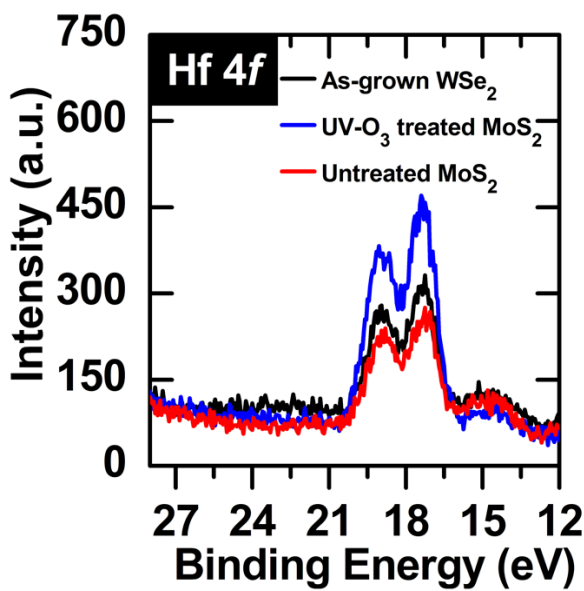

**Figure 2.** Hf 4$f$ spectra after ALD of HZO on as-grown WSe$_2$, UV-O$_3$ treated MoS$_2$, and untreated MoS$_2$.

The untreated MoS$_2$ sample resulted in the lowest Hf 4$f$ peak intensity after HZO deposition. It is likely that the HZO deposited on untreated MoS$_2$ is composed of islands, rather than a uniform HZO film. A poor ALD oxide film morphology on a non-functionalized MoS$_2$ surface has been shown previously in other works [39, 50].

The UV-O$_3$ treated MoS$_2$ sample resulted in the highest Hf 4*f* peak intensity. Prior research has shown that UV-O$_3$ treated MoS$_2$ results in a functionalization of the surface that increases the number of nucleation sites for ALD of oxides on MoS$_2$, which leads to uniform film deposition [39]. The functionalization of the MoS$_2$ surface through UV-O$_3$ exposure, as reported by Azcatl *et al.* [39], is achieved by forming bonds between sulfur and oxygen atoms adsorbed on the surface, without breaking the Mo-S bonds. The oxygen-terminated surface provides ideal nucleation sites for ALD, leading to good film morphology and uniformity. However, long exposures of MoS$_2$ to UV-O$_3$ have also been found to oxidize the Mo, which compromises the MoS$_2$ [37, 54]. In this work, the process developed by Azcatl *et al.* has been found to be transferrable to a bench-top UV lamp, with 30 s of UV-O$_3$ exposure in air. XPS results showing the differences between as-exfoliated (untreated) MoS$_2$ and UV-O$_3$ treated MoS$_2$ can be found in Fig. S1 of Supporting Information. XPS shows that a 30 s UV-O$_3$ treatment provided S-O$_{ads}$ bonds without breaking Mo-S and forming Mo-O.

As shown in Fig. 2, this functionalization allowed improved nucleation and growth of HZO by ALD, as evidenced by the increased Hf 4*f* peak intensity for the UV-O$_3$ treated sample. The ALD was found to be most optimal for UV-O$_3$ treated MoS$_2$, and least optimal for untreated MoS$_2$. The intensities of the Hf 4*f* peaks for the as-grown WSe$_2$ and untreated MoS$_2$ samples are similar, but it will be shown later that the HZO film on untreated MoS$_2$ is highly unstable at elevated temperatures, whereas the HZO film on as-grown WSe$_2$ is comparatively stable. Furthermore, it is known that UV-O$_3$ treatments of WSe$_2$ can lead to etching of the WSe$_2$ [40], so this was not employed here.

HZO Thermal Stability on MoS$_2$

To demonstrate the difference in the thermal stability of HZO on non-functionalized (untreated) vs. functionalized (UV-O$_3$ treated) MoS$_2$, Fig. 3 shows the evolution of the Hf 4$f$ spectra on both types of MoS$_2$ at varying anneal temperatures, conducted under UHV conditions.

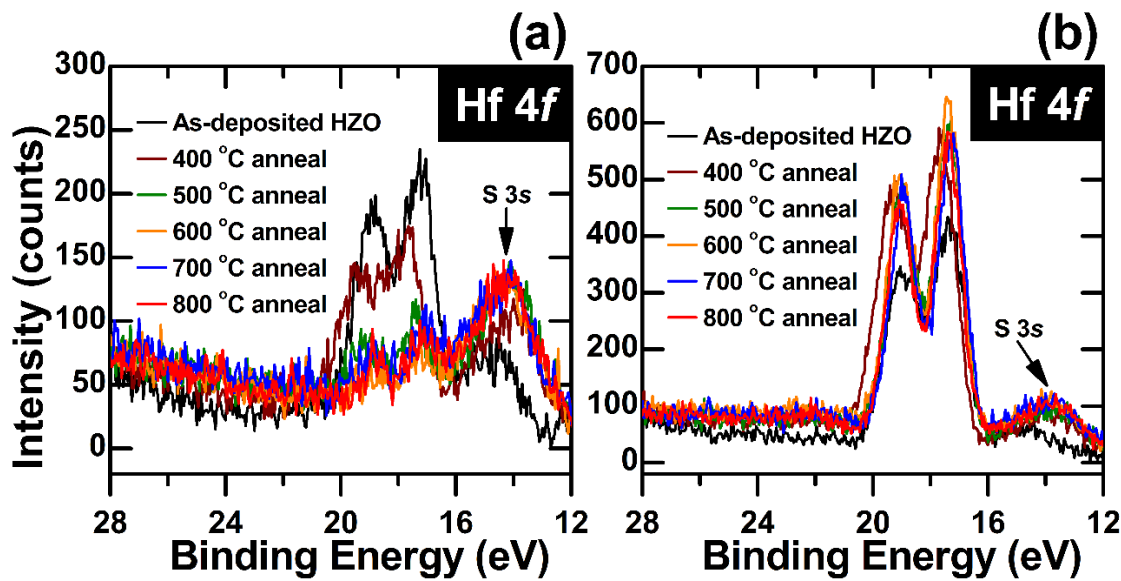

**Figure 3.** Hf 4$f$ spectra of HZO deposited on untreated MoS$_2$, (a), and UV-O$_3$ treated MoS$_2$, (b), after ALD and after each annealing step.

For the non-functionalized MoS$_2$ sample as shown in Fig. 3a, the Hf 4$f$ features at approximately 17 eV and 19 eV decrease in intensity after the 400 °C anneal. At the higher temperature anneals, the Hf 4$f$ peak intensities decrease further. This suggests that the HZO film is removed with heating the untreated MoS$_2$ sample. We acknowledge that a decrease in XPS peak intensity could be interpreted as diffusion of the HZO components into the bulk, but this is unlikely when decreasing peak intensities are not observed for the parallel sample (UV-O$_3$ treated MoS$_2$). The decreasing trend can also be observed with the Zr 3$d$ spectra of the untreated MoS$_2$ sample (shown in Fig. S2 of Supporting Information), and there are no discernable Zr 3$d$ peaks for anneal temperatures higher than 400 °C, further elucidating the removal of the HZO film on untreated MoS$_2$ at high temperatures. Because there are no detectable Zr 3$d$ features remaining at anneal temperatures of 500 °C and higher, while small Hf 4$f$ peaks are still detected, this suggests that the Zr species are desorbed before the Hf species. To verify that this is not an effect of the order in which the ALD precursors were deposited, a separate HZO on untreated MoS$_2$ sample was prepared wherein the deposition order of the precursors was reversed, *i.e.* TEMA-Zr was the first precursor introduced to the growth chamber followed by oxidant and then TEMA-Hf. The results for this additional sample are shown in Supporting Information (Fig. S3) and reveal the same behavior of near complete reduction of Zr 3$d$ signals and significant reduction of Hf 4$f$ peaks. Reasons for preferred sticking of HfO$_2$ over ZrO$_2$ on MoS$_2$ are not clear at this time and will require further investigation. The removal of HZO from the untreated MoS$_2$ surface at temperatures > 400 °C is consistent with the model for ALD on MoS$_2$ proposed in earlier work [50]. In that work, it was suggested that depositions on MoS$_2$ did not occur through reactions with the substrate, but instead relied on reactions between the precursors in physisorbed states on the substrate surface. The reaction products were assumed to be less volatile than the unreacted precursors at the deposition temperature of 200 °C, and it was still assumed to be lacking covalent bonds to the surface as evidenced by the lack of observable changes in the XPS core level spectra of the Mo or S features. This assumption of weakly bonded species that were less volatile than individual precursors would seem to be consistent with our present observations of thermal desorption at higher temperature; however, it cannot exclude that the presence of Zr in the present study may also play a role in the instability.

HZO on a functionalized MoS$_2$ substrate, as shown in Fig. 3b, exhibits a different trend. The initial 400 °C anneal shows an increase in Hf 4*f* intensity compared to the as-deposited curve. This is consistent with adventitious carbon or hydroxide species being desorbed with heating, thus reducing attenuation effects and increasing the signal coming from the HZO film. Note that the desorption of adventitious contaminants also occurs for the non-functionalized MoS$_2$ sample, but the removal of the HZO film dominates over the effects of removing contaminants, with the overall result being the decrease in the Hf 4*f* intensity. It is observed that with increasing the anneal temperature, the Hf 4*f* intensity in Fig. 3b stays consistent, indicating that most, or all adsorbates on top of the HZO film were likely removed from the initial 400 °C anneal. Further, the consistent peak intensity also indicates that the amount of HZO is not reduced with high temperature anneals, suggesting that the HZO on functionalized MoS$_2$ is thermally stable for the full temperature range used in this study. This is contrary to what is observed for untreated MoS$_2$, thus emphasizing that the HZO film deposited on the non-functionalized MoS$_2$ surface (Fig. 3a) exhibits lower thermal stability. This highlights the importance of functionalization for oxide growth on MoS$_2$. As well as exhibiting low coverage, island growth, and a non-ideal interface, as shown previously [39, 50], this present study also demonstrates that ALD of an oxide on a non-functionalized MoS$_2$ surface will produce a thermally unstable film. Other than changes in intensity, it is also notable that the 400 °C curve in Figs. 3a and 3b are shifted to higher binding energy. These peak shifts will be discussed later in the text.

HZO Thermal Stability on WSe$_2$

To examine the thermal stability of the HZO film deposited on MBE-grown WSe$_2$, the evolution of the Hf 4*f* curves after ALD of HZO and after sequential annealing is shown in Fig. 4. The intensities of the Hf features do not change significantly within the temperature range studied. When compared to MoS$_2$, these results demonstrate that HZO deposited on non-functionalized WSe$_2$ has superior thermal stability over a non-functionalized MoS$_2$ substrate.

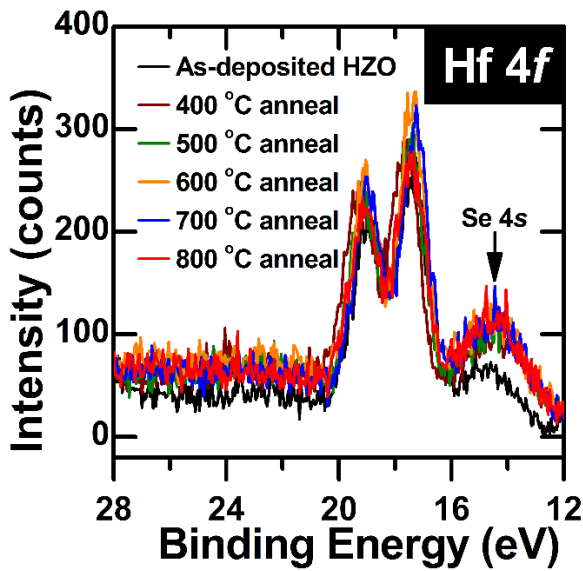

**Figure 4.** Hf 4*f* spectra of HZO deposited on as-grown WSe$_2$ after ALD and after each annealing step.

We now turn our attention to changes in the WSe$_2$ itself at the different annealing temperatures. The normalized W 4*f* spectra after ALD and after each anneal are plotted in Fig. 5a. It is evident that following ALD of HZO, the W is partially oxidized, based on the presence of high binding energy peaks corresponding to WO$_3$ (~35.6 and 37.8 eV). The WO$_3$ peaks are observed to decrease in intensity with increasing anneal temperature, as discussed further later.

The presence of this oxide indicates that excess oxygen in the ALD process, *i.e.* water vapor and impurities in the carrier gas, have oxidized the tungsten. A higher density of defects such as step edges in the as-grown WSe$_2$ sample makes it more susceptible to reaction with oxygen sources in the ALD chamber, thus forming WO$_3$.

Another noteworthy observation from the normalized W 4$f$ spectra is the emergence of a sub-stoichiometric (WSe$_{2-x}$) state after the 800 °C anneal. This state is seen as an asymmetry at the low binding energy side for the 800 °C curve, which has also been observed for MBE growths of WSe$_2$ using low Se:W flux ratios [55]. The spectral deconvolution of the post-800 °C anneal W 4$f$ spectrum is shown in Fig. 5b. In this figure, the low intensity portion of the spectrum is enlarged and emphasized, in order to clearly depict the WSe$_{2-x}$ peaks. The main WSe$_2$ peaks are present at 32.2 eV and 34.4 eV (green peak fits), while the sub-stoichiometric peaks are at 31.5 eV and 33.7 eV (magenta peak fits). The appearance of the sub-stoichiometric state, WSe$_{2-x}$, shows that annealing to temperatures as high as 800 °C volatilizes some of the Se in the WSe$_2$ film, which makes it selenium-deficient and thus may affect its properties. This result shows that our as-grown WSe$_2$ is not thermally stable beyond 700 °C. Similar decomposition of MBE-grown WSe$_2$ above 700 °C has been shown in other work [56]. For the annealing time used in this experiment (20-40 min), we report a thermal budget of 700 °C for the HZO/WSe$_2$ interface. At 700 °C and below, both the HZO and the WSe$_2$ appear to be thermally stable. After annealing this sample to 800 °C, decomposition was observed in the TMDC. Some Se was driven off of the WSe$_2$, leaving behind a sub-stoichiometric component in the film as shown in Fig. 5.

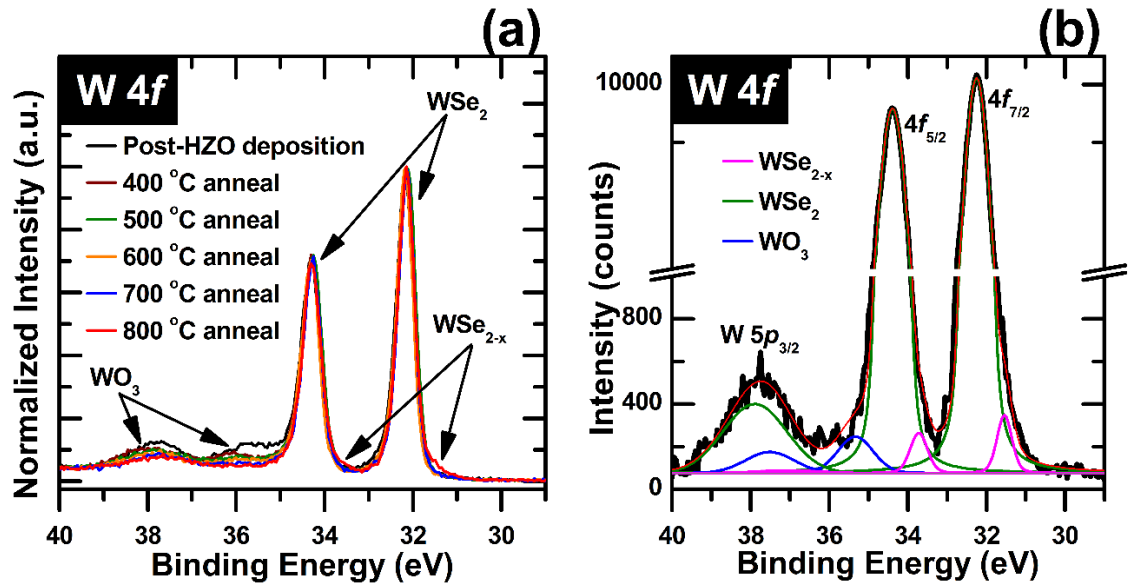

**Figure 5.** (a) Normalized W 4*f* spectra after ALD of HZO and after each annealing step. (b) Spectral deconvolution of 800 °C W 4*f* spectrum (red curve in (a)). Three chemical states are present: $WO_3$, $WSe_2$, and $WSe_{2-x}$. Note that the $WSe_2$ and $WSe_{2-x}$ peaks have similar widths. The $WO_3$ features are broader because the oxide state is expected to be more disordered.

Oxygen Vacancies in HZO

The Hf 4*f* spectra presented above (Figs. 3 and 4) show that the 400 °C curves for all three samples (untreated $MoS_2$, UV-$O_3$ treated $MoS_2$, and as-grown $WSe_2$) are shifted to higher binding energy (+0.3 eV) with respect to their positions from the as-deposited spectra. It should also be noted that at 400 °C, the Zr 3*d* peaks (Fig. S2 of Supporting Information) shift to higher binding energy together with the Hf peaks. Such identical shifts in both Hf and Zr are indicative of an electronic change such as charging or Fermi level shifts. These shifts can be attributed to a loss of oxygen in the HZO with initial heating. Shifts to higher binding energy after one annealing step is consistent with previous observations on $TiO_2$, $HfO_2$, and $ZrO_2$ deposited on Si (100) [57, 58].

It has been reported previously that heating of oxide films causes a loss of some oxygen [57-59], which corresponds to introduction of oxygen vacancies. Oxygen vacancies can serve as *n*-type dopants in $HfO_2$ [60, 61] and $ZrO_2$ [61]. However, after anneals at 500 °C and higher, the Hf 4*f* peaks shift back to lower binding energy, which signifies that the Fermi level moves closer to the valence band, and is thus characteristic of a decrease in *n*-type charge carriers. Therefore, this shift to lower binding energy at the high temperatures is attributed to passivation of oxygen vacancies in the HZO film. This requires a source of atoms responsible for oxygen vacancy passivation, and we propose two possibilities: one using oxygen atoms and the other using chalcogen atoms. We discuss both possibilities in this section of the paper.

We first discuss the proposed vacancy passivation mechanism involving oxygen atoms. Figure 6 shows the O 1*s* spectra of HZO on functionalized $MoS_2$ at different annealing stages. The O 1*s* feature at each step of the process is composed of three different chemical states, as labeled in the figure. The high intensity oxygen peak at the low binding energy side corresponds to metal oxide; the intermediate peak between 532-533 eV is assigned to non-lattice oxygen ($O^{2-}$) present in the material [62-64]; and the high binding energy feature is from hydroxide species adsorbed on the surface, likely from residual $OH^-$ from the ALD process as well as atmospheric exposure after HZO deposition.

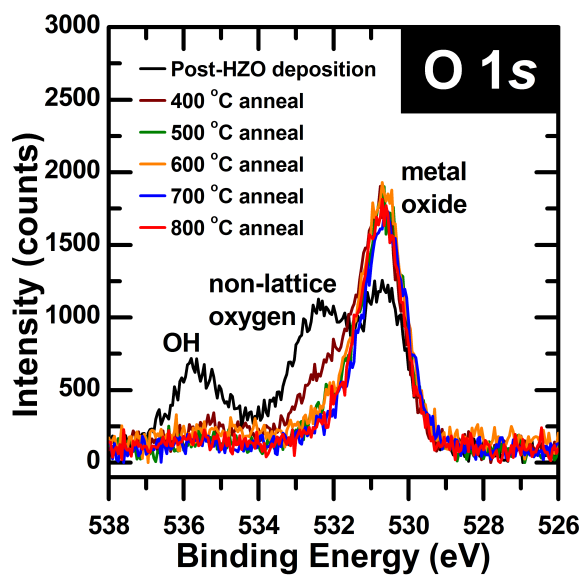

**Figure 6.** (a) O 1*s* spectra of HZO on the functionalized MoS$_2$ sample.

We draw attention to the non-lattice oxygen feature. Here we note that the as-deposited HZO films are amorphous and therefore do not have a lattice arrangement of the atoms. Crystallization of these films is discussed in a later section. In the case that the HZO does not have a lattice, it should be noted that the non-lattice oxygen component corresponds to oxygen atoms that are coordinated differently. For HfO$_2$, the oxygen atoms can have a coordination number of 3 or 4 [65]. We assume non-lattice oxygen to have a non-threefold and non-fourfold coordination. We report that after the HZO deposition, there is an excess of oxygen anions in the material, based on the large non-lattice oxygen peak. Negatively charged excess oxygen after oxide deposition has been previously reported by Fulton *et al.* [57, 58]. Although the majority of this non-lattice oxygen is removed from the film after the first anneal at 400 °C, there is still some O$^{2-}$ remaining after this initial anneal. It is known that non-lattice oxygen can react with oxygen vacancies ultimately passivating the defects and neutralizing charge [62, 63], so it is possible that the non-lattice oxygen is being consumed by the HZO film to passivate oxygen vacancies. We thus speculate that the remaining non-lattice oxygen, or $O^{2-}$ anions, after an initial UHV anneal are more tightly bound to the oxide film, and could act as an oxygen source for passivating oxygen vacancies.

We note that for the HZO on $WSe_2$ sample, the $WO_3$ could also be playing a role in vacancy passivation. As mentioned earlier, it is seen in Fig. 5a that the $WO_3$ features decrease in intensity with increasing temperature. Thus, another potential oxygen source for passivating oxygen vacancies at the higher temperature anneals is this $WO_3$ at the HZO/$WSe_2$ interface. Sowinska *et al.* [64] and Bertaud *et al.* [66] have shown that oxygen vacancies in $HfO_2$ are passivated by a metal oxide ($TiO_y$) interfacial layer between their $HfO_2$ film and a Ti top electrode. In the present study, it is possible that the oxygen from the $WO_3$ is being consumed to passivate oxygen vacancies, which makes the HZO less *n*-type after higher temperature anneals (500-800 °C).

For both HZO on $MoS_2$ and HZO on $WSe_2$, a second potential source for passivating atoms is the TMDC underlayer. Because the HZO is in contact with a TMDC, it cannot be discounted that the chalcogen atoms from the TMDC substrate could be diffusing into the HZO film and passivating any oxygen vacancies in it. Sulfur (from $MoS_2$) and selenium (from $WSe_2$) have the same valence as oxygen, making them both a likely candidate for being a substitutional atom for oxygen in the HZO lattice. For this proposed vacancy passivation mechanism, we speculate that at 400 °C, there is not enough thermal energy for diffusion of excess S or Se from the TMDC into the HZO. This is why the HZO becomes more *n*-type, signifying an increase in oxygen vacancies, at this annealing temperature. The vacancy passivation occurs at higher temperature anneals, wherein any excess chalcogen atoms, such as interstitials or intercalants in the TMDC, are likely to have enough energy to diffuse away from the TMDC substrate and into the HZO. We report that changes in the widths of the XPS features due to this potential phenomenon are not detected in these samples because the concentration of oxygen vacancies is outside the detection limit of XPS.

Based on these data it is not possible to differentiate between oxygen or chalcogen atoms as the source of vacancy passivation. Determining which of these vacancy passivation mechanisms is more likely will be focus of future work.

Crystallization of HZO

Crystallization of the amorphous as-deposited HZO is expected to occur within the temperature range used in this work. With a 10 nm $Hf_{0.5}Zr_{0.5}O_2$ film, Park *et al.* report a mixture of the orthorhombic and tetragonal phases after annealing at 400-500 °C, and a mixture of the monoclinic, orthorhombic, and tetragonal phases after annealing at 600-700 °C. They state that the start of crystallization occurs with the formation of the tetragonal phase, followed by the transformation of some grains into the orthorhombic phase because of a two-dimensional stress effect. At higher temperatures, the monoclinic phase is formed, primarily from the tetragonal phase, as the grain size increases. [32] A study by Hsain *et al.* confirms these phase transitions through in-situ high-temperature XRD, which allowed them to study the phases present in 30 nm films of different HZO compositions as they develop during heating and cooling. Additionally, they revealed that for annealing temperatures of 500-1000 °C, the monoclinic phase fraction increases linearly with increasing temperature. [67]

Using XPS, we have found the HZO to show qualitative signs of ordering after annealing at 500 °C and higher. Figure 7 shows the Hf 4*f* data from Fig. 3b normalized and energy corrected. It is clear from the peak-to-valley ratio, which is the ratio between the maximum intensity and the value between the doublet feature, that the full width at half maximum of the individual components has decreased after annealing at temperatures of 500 °C and higher. This sharpening of the peaks typically indicates some degree of increased ordering of the material. We report that this ordering phenomenon is because of the onset of crystallization of the HZO.

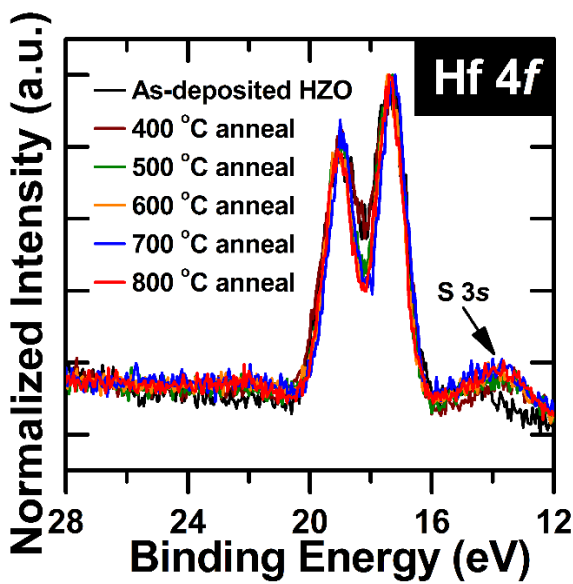

**Figure 7.** Normalized and overlaid Hf $4f$ spectra of HZO on functionalized $MoS_2$ (from Fig. 3b).

All three samples discussed above were for interfacial studies through XPS and thus comprised a very thin layer (~1 nm) of HZO. We fabricated two separate samples for XRD to further examine the crystallization of HZO deposited on top of a TMDC. Approximately 10 layers of $WSe_2$ was MBE-grown on an AlN/Si (111) substrate using the same method described for the $WSe_2$ growth on HOPG. After MBE growth, nominally 20 nm of HZO was deposited on top of the $WSe_2$ through ALD.

The first sample (HZO/WSe$_2$/AlN sample 1) was subjected to a series of anneals in UHV for 40 min at each of the following temperatures: 400 °C, 500 °C, and 600 °C. This annealing procedure replicates that of the samples in the XPS study. GIXRD measurements were performed after each annealing step. No peaks were observed in any of the diffraction measurements; the pattern measured after the 600 °C anneal is shown in blue in Fig. 8.

The second sample (HZO/WSe$_2$/AlN sample 2) was subjected to a 600 °C anneal in UHV for 80 min rather than 40 min to allow more time for grain growth. The GIXRD pattern for this sample after the 600 °C anneal is shown in black in Fig. 8. The pattern indicates that the HZO has crystallized and demonstrates mixed phase assemblage due to the presence of both monoclinic and orthorhombic and/or tetragonal peaks. The peak at 30.5° presents challenges in indexing due to similar *d*-spacings of the orthorhombic and tetragonal phases; therefore, this peak could be the result of the orthorhombic (111) reflection, the tetragonal (101) reflection, or both. The peak at 31.7° can be attributed to the monoclinic (111) reflection. It is interesting to note that the monoclinic ($\bar{1}$11) reflection at 28.3° is of much lower intensity than the monoclinic (111) reflection, suggesting that some degree of crystallographic texture exists, or that the strain state of the film has resulted in preferred orientation of the monoclinic phase, as the (111) and ($\bar{1}$11) reflections represent planes of the same family that are related by a ferroelastic transformation.

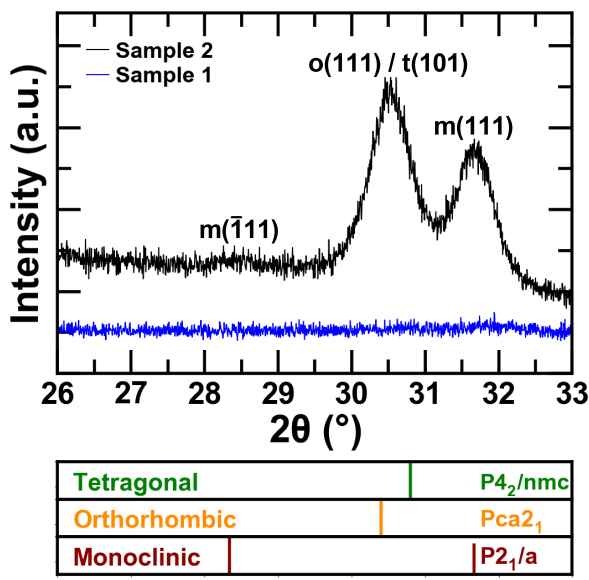

**Figure 8**. GIXRD patterns of the 20 nm thick HZO on WSe$_2$ samples. For these measurements, the X-ray incident angle was fixed at 0.7°, and the 2θ range of 26-33° was measured. The pattern for sample 1 after the 600 °C annealing step for 40 min is shown in blue, while the pattern for sample 2 after the 600 °C anneal for 80 min is shown in black. Markers designate reflections for tetragonal, orthorhombic, and monoclinic phases.

We note that while a 40 min anneal, used for HZO/WSe$_2$/AlN sample 1, was sufficient to observe peak narrowing by XPS, it did not yield detectable peaks in XRD. We speculate that 40 min was sufficient to generate short range atomic order, possibly in the near-surface region to which XPS is most sensitive, but was insufficient to suitably coarsen them. In previous work it was shown that newly formed grains in 2.8 nm thick HfO$_2$ were ~1-2 nm in size [59]. These grain sizes would be expected to produce peaks that are too broad (~10 degrees in 2θ) to be observable by XRD. For HZO/WSe$_2$/AlN sample 2, the anneal time was longer to allow grain coarsening, such that XRD peaks are observable (Fig. 8).

Conclusions

We have examined the thermal stability of HZO grown directly atop TMDCs (MoS$_2$ and WSe$_2$) of interest for next generation non-volatile memories and advanced low power transistors. For MoS$_2$, the importance of functionalizing the MoS$_2$ prior to ALD, in order to produce a high-quality and thermally stable oxide film, was demonstrated. The HZO film deposited on a non-functionalized MoS$_2$ substrate was unstable with temperature, wherein removal of the HZO overlayer began at a 400 °C anneal. ZrO$_2$ species were fully removed after a 500 °C anneal, leaving behind trace amounts of hafnium oxide, which shows that HfO$_2$ sticking on the MoS$_2$ surface is preferred over the sticking of ZrO$_2$. HZO on a functionalized UV-O$_3$ treated MoS$_2$ substrate, on the other hand, did not show signs of decomposition, and both the HZO and the MoS$_2$ were stable between 400-800 °C. For the as-grown WSe$_2$ film, a thermal budget of 700 °C is reported. The W 4*f* spectra showed an asymmetry in the low binding energy side after an 800 °C anneal, which signifies the appearance of a sub-stoichiometric state, WSe$_{2-x}$. We also report on possible mechanisms for passivating oxygen vacancies in the HZO. Lastly, crystallization of the HZO was examined, and the monoclinic, tetragonal, and orthorhombic phases were observed via XRD after annealing for 80 min at 600 °C, demonstrating that functional crystalline films can be directly integrated on TMDCs.

Acknowledgments

The authors acknowledge the Research Innovation Award program within the School of Engineering and Applied Science at the University of Virginia for support of this work. M.G.S. acknowledges support through the William L. Ballard Jr. Endowed Graduate Fellowship. S.T.J.,

S.S.F., and J.F.I. acknowledge the Semiconductor Research Corporation's Nanomanufacturing Materials and Processing program for support under task number 2875.001.References

[1] J.F. Scott, C.A. Paz de Araujo, Ferroelectric Memories, Science, 246 (1989) 1400.
[2] J.F. Ihlefeld, D.T. Harris, R. Keech, J.L. Jones, J.-P. Maria, S. Trolier-McKinstry, Scaling Effects in Perovskite Ferroelectrics: Fundamental Limits and Process-Structure-Property Relations, Journal of the American Ceramic Society, 99 (2016) 2537-2557.
[3] K.R. Udayakumar, T.S. Moise, S.R. Summerfelt, K. Boku, K. Remack, J. Rodriguez, M. Arendt, G. Shinn, J. Eliason, R. Bailey, P. Staubs, Manufacturable High-Density 8 Mbit One Transistor–One Capacitor Embedded Ferroelectric Random Access Memory, Japanese Journal of Applied Physics, 47 (2008) 2710-2713.
[4] T.S. Böscke, J. Müller, D. Bräuhaus, U. Schröder, U. Böttger, Ferroelectricity in hafnium oxide thin films, Applied Physics Letters, 99 (2011) 102903.
[5] S. Mueller, S.R. Summerfelt, J. Muller, U. Schroeder, T. Mikolajick, Ten-Nanometer Ferroelectric Si:$HfO_2$ Films for Next-Generation FRAM Capacitors, IEEE Electron Device Letters, 33 (2012) 1300-1302.
[6] H. Mulaosmanovic, J. Ocker, S. Müller, U. Schroeder, J. Müller, P. Polakowski, S. Flachowsky, R. van Bentum, T. Mikolajick, S. Slesazeck, Switching Kinetics in Nanoscale Hafnium Oxide Based Ferroelectric Field-Effect Transistors, ACS Applied Materials & Interfaces, 9 (2017) 3792-3798.
[7] H. Mulaosmanovic, T. Mikolajick, S. Slesazeck, Accumulative Polarization Reversal in Nanoscale Ferroelectric Transistors, ACS Applied Materials & Interfaces, 10 (2018) 23997-24002.
[8] F.A. McGuire, Y.-C. Lin, K. Price, G.B. Rayner, S. Khandelwal, S. Salahuddin, A.D. Franklin, Sustained Sub-60 mV/decade Switching via the Negative Capacitance Effect in $MoS_2$ Transistors, Nano Letters, 17 (2017) 4801-4806.
[9] M. Si, C. Jiang, W. Chung, Y. Du, M.A. Alam, P.D. Ye, Steep-Slope $WSe_2$ Negative Capacitance Field-Effect Transistor, Nano Letters, 18 (2018) 3682-3687.
[10] X. Sang, E.D. Grimley, T. Schenk, U. Schroeder, J.M. LeBeau, On the structural origins of ferroelectricity in $HfO_2$ thin films, Applied Physics Letters, 106 (2015) 162905.
[11] H.J. Kim, M.H. Park, Y.J. Kim, Y.H. Lee, W. Jeon, T. Gwon, T. Moon, K.D. Kim, C.S. Hwang, Grain size engineering for ferroelectric $Hf_{0.5}Zr_{0.5}O_2$ films by an insertion of $Al_2O_3$ interlayer, Applied Physics Letters, 105 (2014) 192903.
[12] C. Künneth, R. Materlik, A. Kersch, Modeling ferroelectric film properties and size effects from tetragonal interlayer in $Hf_{1-x}Zr_xO_2$ grains, Journal of Applied Physics, 121 (2017) 205304.
[13] T. Shiraishi, K. Katayama, T. Yokouchi, T. Shimizu, T. Oikawa, O. Sakata, H. Uchida, Y. Imai, T. Kiguchi, T.J. Konno, H. Funakubo, Impact of mechanical stress on ferroelectricity in $(Hf_{0.5}Zr_{0.5})O_2$ thin films, Applied Physics Letters, 108 (2016) 262904.
[14] S.J. Kim, D. Narayan, J.-G. Lee, J. Mohan, J.S. Lee, J. Lee, H.S. Kim, Y.-C. Byun, A.T. Lucero, C.D. Young, S.R. Summerfelt, T. San, L. Colombo, J. Kim, Large ferroelectric polarization of TiN/$Hf_{0.5}Zr_{0.5}O_2$/TiN capacitors due to stress-induced crystallization at low thermal budget, Applied Physics Letters, 111 (2017) 242901.
[15] S.J. Kim, J. Mohan, J.S. Lee, H.S. Kim, J. Lee, C.D. Young, L. Colombo, S.R. Summerfelt, T. San, J. Kim, Stress-Induced Crystallization of Thin $Hf_{1-x}Zr_xO_2$ Films: The Origin of Enhanced26

Energy Density with Minimized Energy Loss for Lead-Free Electrostatic Energy Storage Applications, ACS Applied Materials & Interfaces, 11 (2019) 5208-5214.
[16] E.D. Grimley, T. Schenk, X. Sang, M. Pešić, U. Schroeder, T. Mikolajick, J.M. LeBeau, Structural Changes Underlying Field-Cycling Phenomena in Ferroelectric $HfO_2$ Thin Films, Advanced Electronic Materials, 2 (2016) 1600173.
[17] S. Starschich, S. Menzel, U. Böttger, Pulse wake-up and breakdown investigation of ferroelectric yttrium doped $HfO_2$, Journal of Applied Physics, 121 (2017) 154102.
[18] S. Mueller, J. Mueller, A. Singh, S. Riedel, J. Sundqvist, U. Schroeder, T. Mikolajick, Incipient Ferroelectricity in Al-Doped $HfO_2$ Thin Films, Advanced Functional Materials, 22 (2012) 2412-2417.
[19] J. Müller, T.S. Böscke, U. Schröder, S. Mueller, D. Bräuhaus, U. Böttger, L. Frey, T. Mikolajick, Ferroelectricity in Simple Binary $ZrO_2$ and $HfO_2$, Nano Letters, 12 (2012) 4318-4323.
[20] J. Müller, U. Schröder, T.S. Böscke, I. Müller, U. Böttger, L. Wilde, J. Sundqvist, M. Lemberger, P. Kücher, T. Mikolajick, L. Frey, Ferroelectricity in yttrium-doped hafnium oxide, Journal of Applied Physics, 110 (2011) 114113.
[21] S. Mueller, C. Adelmann, A. Singh, S. Van Elshocht, U. Schroeder, T. Mikolajick, Ferroelectricity in Gd-Doped $HfO_2$ Thin Films, ECS Journal of Solid State Science and Technology, 1 (2012) N123-N126.
[22] H. Yu, C.-C. Chung, N. Shewmon, S. Ho, J.H. Carpenter, R. Larrabee, T. Sun, J.L. Jones, H. Ade, B.T. O'Connor, F. So, Flexible Inorganic Ferroelectric Thin Films for Nonvolatile Memory Devices, Advanced Functional Materials, 27 (2017) 1700461.
[23] M. Hyuk Park, H. Joon Kim, Y. Jin Kim, W. Lee, H. Kyeom Kim, C. Seong Hwang, Effect of forming gas annealing on the ferroelectric properties of $Hf_{0.5}Zr_{0.5}O_2$ thin films with and without Pt electrodes, Applied Physics Letters, 102 (2013) 112914.
[24] A. Lipatov, P. Sharma, A. Gruverman, A. Sinitskii, Optoelectrical Molybdenum Disulfide ($MoS_2$)—Ferroelectric Memories, ACS Nano, 9 (2015) 8089-8098.
[25] C.H. Li, K.M. McCreary, B.T. Jonker, Spatial Control of Photoluminescence at Room Temperature by Ferroelectric Domains in Monolayer $WS_2$/PZT Hybrid Structures, ACS Omega, 1 (2016) 1075-1080.
[26] J.P.B. Silva, C. Almeida Marques, J.A. Moreira, O. Conde, Resistive switching in $MoSe_2$/$BaTiO_3$ hybrid structures, Journal of Materials Chemistry C, 5 (2017) 10353-10359.
[27] H.S. Lee, S.-W. Min, M.K. Park, Y.T. Lee, P.J. Jeon, J.H. Kim, S. Ryu, S. Im, $MoS_2$ Nanosheets for Top-Gate Nonvolatile Memory Transistor Channel, Small, 8 (2012) 3111-3115.
[28] E. Preciado, F.J.R. Schülein, A.E. Nguyen, D. Barroso, M. Isarraraz, G. von Son, I.H. Lu, W. Michailow, B. Möller, V. Klee, J. Mann, A. Wixforth, L. Bartels, H.J. Krenner, Scalable fabrication of a hybrid field-effect and acousto-electric device by direct growth of monolayer $MoS_2$/$LiNbO_3$, Nature Communications, 6 (2015) 8593.
[29] I.P. Batra, P. Wurfel, B.D. Silverman, Depolarization Field and Stability Considerations in Thin Ferroelectric Films, Journal of Vacuum Science and Technology, 10 (1973) 687-692.
[30] H. Ishiwara, Recent progress of FET-type ferroelectric memories, Integrated Ferroelectrics, 34 (2001) 11-20.
[31] H. Ishiwara, Current status and prospects of FET-type ferroelectric memories, Journal of Semiconductor Technology and Science, 1 (2001) 1-14.
[32] M. Hyuk Park, H. Joon Kim, Y. Jin Kim, W. Lee, T. Moon, C. Seong Hwang, Evolution of phases and ferroelectric properties of thin $Hf_{0.5}Zr_{0.5}O_2$ films according to the thickness and annealing temperature, Applied Physics Letters, 102 (2013) 242905.
[33] D.S. Macintyre, O. Ignatova, S. Thoms, I.G. Thayne, Resist residues and transistor gate fabrication, Journal of Vacuum Science & Technology B: Microelectronics and Nanometer Structures Processing, Measurement, and Phenomena, 27 (2009) 2597-2601.
[34] A. Pirkle, J. Chan, A. Venugopal, D. Hinojos, C.W. Magnuson, S. McDonnell, L. Colombo, E.M. Vogel, R.S. Ruoff, R.M. Wallace, The effect of chemical residues on the physical and electrical properties of chemical vapor deposited graphene transferred to $SiO_2$, Applied Physics Letters, 99 (2011) 122108.

[35] Y.-C. Lin, C.-C. Lu, C.-H. Yeh, C. Jin, K. Suenaga, P.-W. Chiu, Graphene Annealing: How Clean Can It Be?, Nano Letters, 12 (2012) 414-419.
[36] R. Li, Z. Li, E. Pambou, P. Gutfreund, T.A. Waigh, J.R.P. Webster, J.R. Lu, Determination of PMMA Residues on a Chemical-Vapor-Deposited Monolayer of Graphene by Neutron Reflection and Atomic Force Microscopy, Langmuir, 34 (2018) 1827-1833.
[37] K.M. Freedy, M.G. Sales, P.M. Litwin, S. Krylyuk, P. Mohapatra, A. Ismach, A.V. Davydov, S.J. McDonnell, $MoS_2$ cleaning by acetone and UV-ozone: Geological and synthetic material, Applied Surface Science, 478 (2019) 183-188.
[38] https://www.2spi.com/
[39] A. Azcatl, S. McDonnell, S. K. C, X. Peng, H. Dong, X. Qin, R. Addou, G.I. Mordi, N. Lu, J. Kim, M.J. Kim, K. Cho, R.M. Wallace, $MoS_2$ functionalization for ultra-thin atomic layer deposited dielectrics, Applied Physics Letters, 104 (2014) 111601.
[40] A. Azcatl, S. Kc, X. Peng, N. Lu, S. McDonnell, X. Qin, F. de Dios, R. Addou, J. Kim, M.J. Kim, K. Cho, R.M. Wallace, $HfO_2$ on UV–$O_3$ exposed transition metal dichalcogenides: interfacial reactions study, 2D Materials, 2 (2015) 014004.
[41] P. Zhao, P.B. Vyas, S. McDonnell, P. Bolshakov-Barrett, A. Azcatl, C.L. Hinkle, P.K. Hurley, R.M. Wallace, C.D. Young, Electrical characterization of top-gated molybdenum disulfide metal–oxide–semiconductor capacitors with high-k dielectrics, Microelectronic Engineering, 147 (2015) 151-154.
[42] P. Bolshakov, P. Zhao, A. Azcatl, P.K. Hurley, R.M. Wallace, C.D. Young, Electrical characterization of top-gated molybdenum disulfide field-effect-transistors with high-k dielectrics, Microelectronic Engineering, 178 (2017) 190-193.
[43] P. Zhao, A. Azcatl, Y.Y. Gomeniuk, P. Bolshakov, M. Schmidt, S.J. McDonnell, C.L. Hinkle, P.K. Hurley, R.M. Wallace, C.D. Young, Probing Interface Defects in Top-Gated $MoS_2$ Transistors with Impedance Spectroscopy, ACS Applied Materials & Interfaces, 9 (2017) 24348-24356.
[44] K.M. Freedy, P.M. Litwin, S.J. McDonnell, (Invited) In-Vacuo Studies of Transition Metal Dichalcogenide Synthesis and Layered Material Integration, ECS Transactions, 77 (2017) 11-25.
[45] H. Yamamoto, K. Yoshii, K. Saiki, A. Koma, Improved heteroepitaxial growth of layered $NbSe_2$ on GaAs (111)B, Journal of Vacuum Science & Technology A, 12 (1994) 125-129.
[46] L.A. Walsh, R. Yue, Q. Wang, A.T. Barton, R. Addou, C.M. Smyth, H. Zhu, J. Kim, L. Colombo, M.J. Kim, R.M. Wallace, C.L. Hinkle, $WTe_2$ thin films grown by beam-interrupted molecular beam epitaxy, 2D Materials, 4 (2017) 025044.
[47] M. Nakano, Y. Wang, Y. Kashiwabara, H. Matsuoka, Y. Iwasa, Layer-by-Layer Epitaxial Growth of Scalable $WSe_2$ on Sapphire by Molecular Beam Epitaxy, Nano Letters, 17 (2017) 5595-5599.
[48] X. Zhang, T.H. Choudhury, M. Chubarov, Y. Xiang, B. Jariwala, F. Zhang, N. Alem, G.-C. Wang, J.A. Robinson, J.M. Redwing, Diffusion-Controlled Epitaxy of Large Area Coalesced $WSe_2$ Monolayers on Sapphire, Nano Letters, 18 (2018) 1049-1056.
[49] http://www.kolibrik.net/
[50] S. McDonnell, B. Brennan, A. Azcatl, N. Lu, H. Dong, C. Buie, J. Kim, C.L. Hinkle, M.J. Kim, R.M. Wallace, $HfO_2$ on $MoS_2$ by Atomic Layer Deposition: Adsorption Mechanisms and Thickness Scalability, ACS Nano, 7 (2013) 10354-10361.
[51] V.V. Afanas'ev, D. Chiappe, C. Huyghebaert, I. Radu, S. De Gendt, M. Houssa, A. Stesmans, Band alignment at interfaces of few-monolayer $MoS_2$ with $SiO_2$ and $HfO_2$, Microelectronic Engineering, 147 (2015) 294-297.
[52] K.M. Price, A.D. Franklin, Integration of 3.4 nm $HfO_2$ into the gate stack of $MoS_2$ and $WSe_2$ top-gate field-effect transistors, in: 2017 75th Annual Device Research Conference (DRC), 2017, pp. 1-2.
[53] S.K. Pradhan, B. Xiao, A.K. Pradhan, Energy band alignment of high-k oxide heterostructures at $MoS_2/Al_2O_3$ and $MoS_2/ZrO_2$ interfaces, Journal of Applied Physics, 120 (2016) 125305.
[54] S.J. McDonnell, R.M. Wallace, UV-Ozone Functionalization of 2D Materials, JOM, 71 (2019) 224-237.

[55] P.M. Litwin, M.G. Sales, V. Nilsson, P.V. Balachandran, C. Constantin, S. McDonnell, The effect of growth temperature and metal-to-chalcogen on the growth of WSe2 by molecular beam epitaxy, SPIE, 2019.
[56] W. Chen, X. Xie, J. Zong, T. Chen, D. Lin, F. Yu, S. Jin, L. Zhou, J. Zou, J. Sun, X. Xi, Y. Zhang, Growth and Thermo-driven Crystalline Phase Transition of Metastable Monolayer 1T′-$WSe_2$ Thin Film, Scientific Reports, 9 (2019) 2685.
[57] C.C. Fulton, G. Lucovsky, R.J. Nemanich, Process-dependent band structure changes of transition-metal (Ti,Zr,Hf) oxides on Si (100), Applied Physics Letters, 84 (2004) 580-582.
[58] C.C. Fulton, T.E. Cook, G. Lucovsky, R.J. Nemanich, Interface instabilities and electronic properties of $ZrO_2$ on silicon (100), Journal of Applied Physics, 96 (2004) 2665-2673.
[59] D. Lim, R. Haight, Temperature dependent defect formation and charging in hafnium oxides and silicates, Journal of Vacuum Science & Technology B: Microelectronics and Nanometer Structures Processing, Measurement, and Phenomena, 23 (2005) 201-205.
[60] K. Xiong, J. Robertson, M.C. Gibson, S.J. Clark, Defect energy levels in $HfO_2$ high-dielectric-constant gate oxide, Applied Physics Letters, 87 (2005) 183505.
[61] J. Robertson, K. Xiong, S.J. Clark, Band gaps and defect levels in functional oxides, Thin Solid Films, 496 (2006) 1-7.
[62] M.K. Yang, G.H. Kim, H. Ju, J.-K. Lee, H.-C. Ryu, An analysis of "non-lattice" oxygen concentration effect on electrical endurance characteristic in resistive switching $MnO_x$ thin film, Applied Physics Letters, 106 (2015) 053504.
[63] M.K. Yang, J.-W. Park, T.K. Ko, J.-K. Lee, Bipolar resistive switching behavior in Ti/$MnO_2$/Pt structure for nonvolatile memory devices, Applied Physics Letters, 95 (2009) 042105.
[64] M. Sowinska, T. Bertaud, D. Walczyk, S. Thiess, M.A. Schubert, M. Lukosius, W. Drube, C. Walczyk, T. Schroeder, Hard x-ray photoelectron spectroscopy study of the electroforming in Ti/$HfO_2$-based resistive switching structures, Applied Physics Letters, 100 (2012) 233509.
[65] M.H. Park, T. Schenk, C.M. Fancher, E.D. Grimley, C. Zhou, C. Richter, J.M. LeBeau, J.L. Jones, T. Mikolajick, U. Schroeder, A comprehensive study on the structural evolution of $HfO_2$ thin films doped with various dopants, Journal of Materials Chemistry C, 5 (2017) 4677-4690.
[66] T. Bertaud, M. Sowinska, D. Walczyk, S. Thiess, A. Gloskovskii, C. Walczyk, T. Schroeder, In-operando and non-destructive analysis of the resistive switching in the Ti/$HfO_2$/TiN-based system by hard x-ray photoelectron spectroscopy, Applied Physics Letters, 101 (2012) 143501.
[67] H.A. Hsain, Y. Lee, G. Parsons, J.L. Jones, Compositional dependence of crystallization temperatures and phase evolution in hafnia-zirconia ($Hf_xZr_{1-x}$)$O_2$ thin films, Applied Physics Letters, 116 (2020) 192901.

# Supporting Information

# Thermal Stability of Hafnium Zirconium Oxide on Transition Metal Dichalcogenides

*Maria Gabriela Sales[1], Samantha T. Jaszewski[1], Shelby S. Fields[1], Jon F. Ihlefeld[1,2], Stephen J. McDonnell[1,*]*

[1]Department of Materials Science and Engineering, University of Virginia, Charlottesville, VA 22904

[2]Department of Electrical and Computer Engineering, University of Virginia, Charlottesville, VA 22904

*Corresponding author: mcdonnell@virginia.edu

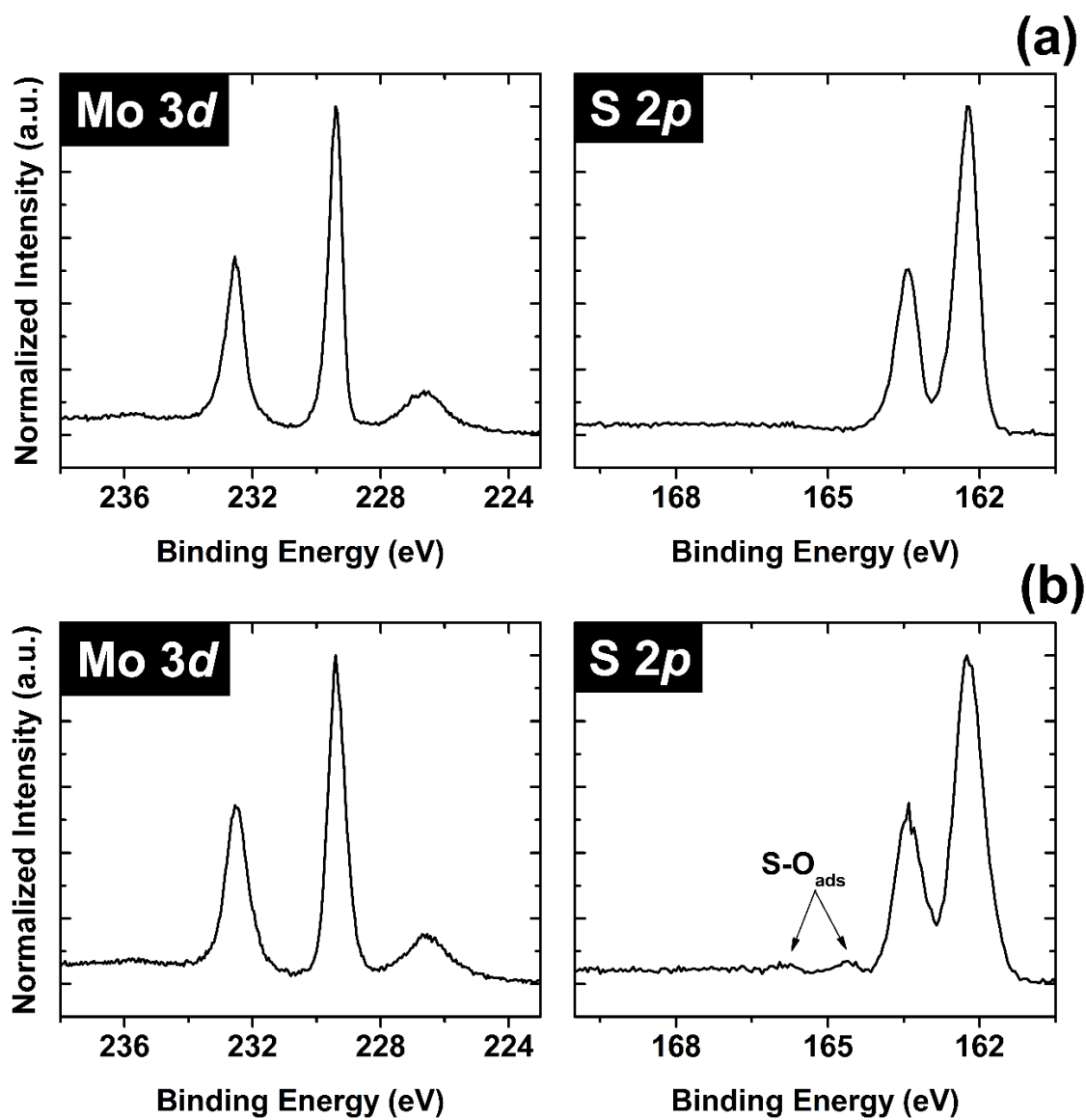

**Figure S1.** Mo 3$d$ and S 2$p$ spectra of MoS$_2$, as-exfoliated (untreated) (a) and after UV-O$_3$ exposure in air for 30 s (b). Through UV-O$_3$ treatment, the MoS$_2$ is functionalized by adsorbed oxygen atoms on the surface, forming S-O$_{ads}$ bonds that are confirmed with XPS.

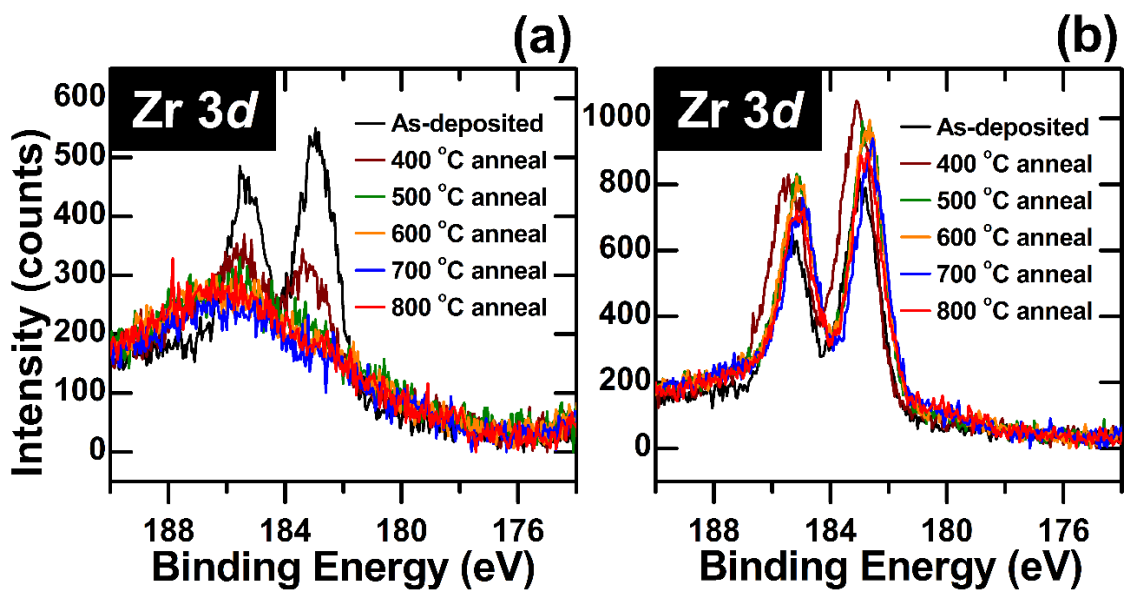

**Figure S2.** Zr 3*d* spectra of HZO deposited on untreated MoS$_2$, (a), and UV-O$_3$ treated MoS$_2$, (b), after ALD and after each annealing step. For untreated MoS$_2$, the Zr 3*d* peaks significantly decrease in intensity after annealing, and are no longer discernable after annealing to 500 °C and higher. For the spectra of both samples, note that the peaks for the 400 °C anneal curves (maroon) are shifted to higher binding energy (~0.3 eV higher), exhibiting a similar trend as the Hf 4*f* spectra (Figs. 3 and 4). This shift to higher binding energy after the first annealing step is attributed to a loss of oxygen in the HZO with initial heating.

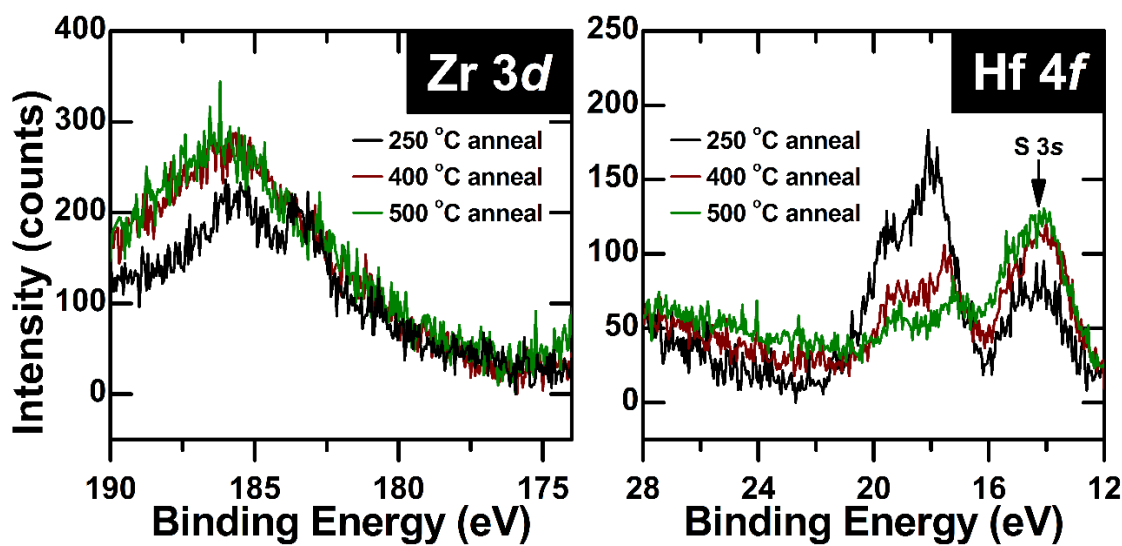

**Figure S3.** Zr 3*d* and Hf 4*f* spectra of an additional HZO on untreated MoS$_2$ sample. For this sample, the Zr precursor was deposited before the Hf precursor, which is the opposite deposition order for the HZO samples discussed in the main paper. It can be seen that even with the order of the precursors reversed, the Hf species are still retained on the sample surface better than the Zr species. The Zr 3*d* doublet (seen at ~183 eV and ~186 eV in the black curve) is not present after the 400 °C anneal, indicating loss of the Zr species, but detectable peaks are still evident in the Hf 4*f* spectra (~17 eV and ~19 eV), which are indicative of Hf species remaining on the sample.